# The socio–economic impact of a breakthrough in the particle accelerators' technology: a research agenda

Massimo Florio

University of Milan

Andrea Bastianin

University of Milan

Paolo Castelnovo

University of Milan

January, 2018

**Abstract:** Preliminary evidence on the long–run trajectory of the accelerator industry suggests that it may be close to the maturity phase of its cycle. If this is the case, how can we measure the benefits of an uncertain breakthrough in acceleration technology? Who are the main stakeholders interested by such a breakthrough? We identify these subjects and sketch some avenues for answering these questions. We thus present a model for the social Cost-Benefit Analysis (CBA) of research infrastructures and illustrate the results of its implementation for assessing the benefits of accelerators in basic science and hadrontherapy. Lastly, we move from the social CBA of single research infrastructures to modeling a major change in the accelerator technology and hence in the industry. A research agenda on the potential impacts of a technological breakthrough is presented.

*Corresponding author*: Massimo Florio, University of Milan, Department of Economics, Management and Quantitative Methods,Via Conservatorio 7, 20122 Milan, Italy. Email: massimo.florio@unimi.it.

# 1 Introduction

The technology underlying particle accelerators has been evolving for more than a century since the first experiments at the Cavendish Laboratory in Cambridge [32]. Through time accelerators have found several applications beyond basic research, from medical to industrial uses. Nowadays, probably more than 40,000 accelerators are operated [7], from the Mev– to the Tev–energy scale, from the meter to the kilometer-length, and from the thousands to the billions Euro cost range.

While new giant machines based on the evolution of existing technology are under study [39], such as the Future Circular Collider [22] or the International Linear Collider [26], new concepts are explored for smaller and less costly machines. One example is plasma wakefield acceleration, with experimentation currently going on in several major laboratories, such as those involved in the EuPRAXIA project [15], in the AWAKE collaboration at CERN [2, 5] and in different projects in the US [8]. There are several other competing ideas for advanced accelerators and it is still uncertain when and which of these new technologies will become available. In any case, such technological breakthrough would probably have the feature of radical innovation, typically associated to contagion effects in different fields of application [1, 11].

Our research questions are the following: (*i*)"*how can we measure the benefits of an uncertain breakthrough innovation in acceleration technology?*" (*ii*) "*Who are the social stakeholders of such an innovation?*". We show that the accelerator industry might have entered into the maturity phase of its cycle; we thus sketch a methodological approach and a research agenda for assessing the benefits and identifying the stakeholders of a technological breakthrough in the "accelerator business". The structure of the paper is as follows: Section 2 reviews some evidence on the trajectory of the accelerator industry; Section 3 presents a simple social Cost-Benefit Analysis (CBA) model and the results of its application to the analysis of accelerators in science and medicine; Section 4 discusses complications posed by shifting from the social CBA of individual accelerators to that of a major change in the technology and in the industry; Section 5 proposes a research agenda and concludes.

# 2 The accelerator industry trajectory

[7] review the literature on the accelerator business and based on a meta–data analysis, forecast that as of 2014 there were 42,200 accelerators worldwide: 27,000 (64%) in industry, 14,000 (33%) for medical purposes and 1,200 (3%) for basic research. These figures exclude electron microscopes and $x$–ray tubes, and the security and defense industries. Accelerators for non–destructive testing and inspection produced between 1950 and 2010 are estimated to be over 1500. About 50% of these are used for security and cargo inspection that represents a fast–growing segment of the accelerator market [25] because of increasing concerns related to illegal activities. [33] estimates that as of 2010, with about 200–300 accelerators produced yearly, this segment is worth US$250M.

[25] report that over 70 companies and institutes produce accelerators for industrial applications; these organizations sell more than 1,100 industrial systems per year — almost twice the number produced for research

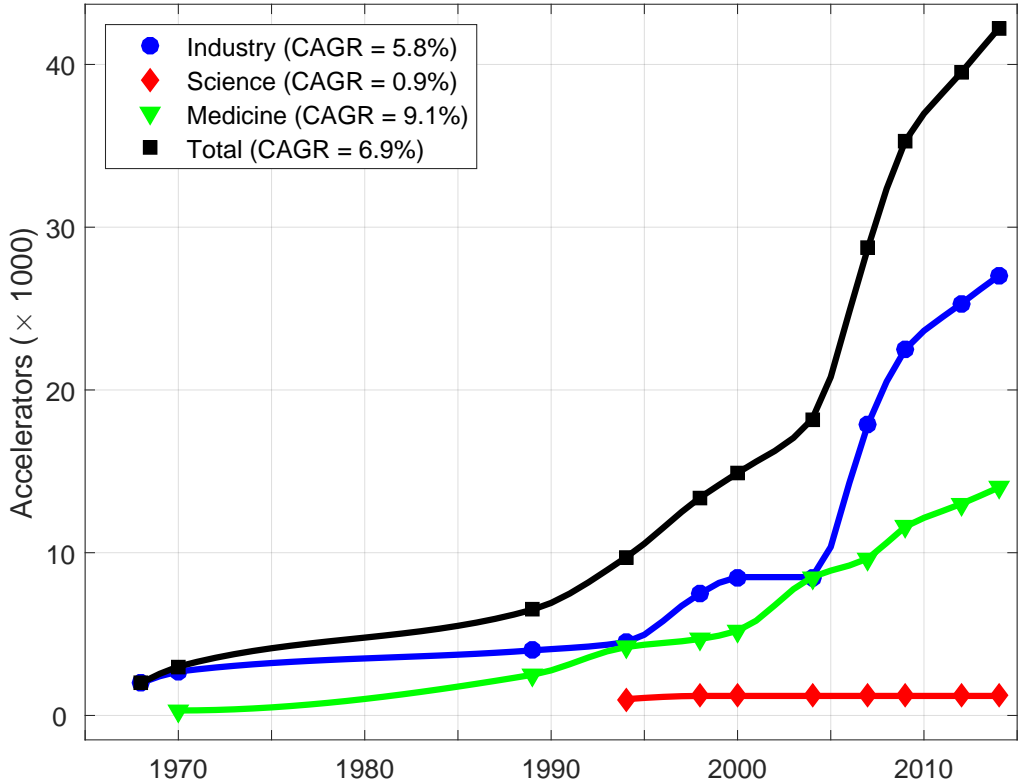

Figure 1: Number of accelerators in the world: total and by field of application, 1968-2014

*Notes:* authors' calculation based on data from Table 1 in [7]. Lines represent the cubic spline interpolation of the original data that are provided only for selected years denoted with symbols. The legend of each graph shows the Compound Annual Growth Rate ($CAGR$) calculated as: $CAGR = (X_T/X_0)^{1/T} - 1$.

or medical therapy — at a market value of $2.2B. Over $1B of this amount is generated by the sales of accelerators for ion implantation into materials — primarily semiconductor devices — whose worldwide value of production is about $300B (see also [38]).

There is some information that can be used to produce scenarios about the future evolution of the accelerator industry. Epidemiological studies can be exploited to extrapolate long–term trends of accelerator sales to hospitals. Socio–demographic factors are another key information: in fact, in low– and middle–income countries the access to radiotherapy is currently limited, hence the income growth in these countries will boost the demand of accelerators for medicine. The decay of the existing stock of accelerators is an additional demand driver, given the heterogeneity in the life–cycles of the machines currently in use. For instance, while in the EU most machines used for radiotherapy are modern linacs, in Eastern Europe and Asia two thirds of the machines are older Cobalt–60 units that will be probably replaced in the future.[1] Similarly, new scenarios for industrial applications — such as ion implantation in the semiconductor industry — may lead to demand forecasts.

---

[1] Up–to–date information about particle accelerators in medicine can be found in the Directory of Radiotherapy Centres (DIRAC) [14] register maintained by International Atomic Energy Agency.

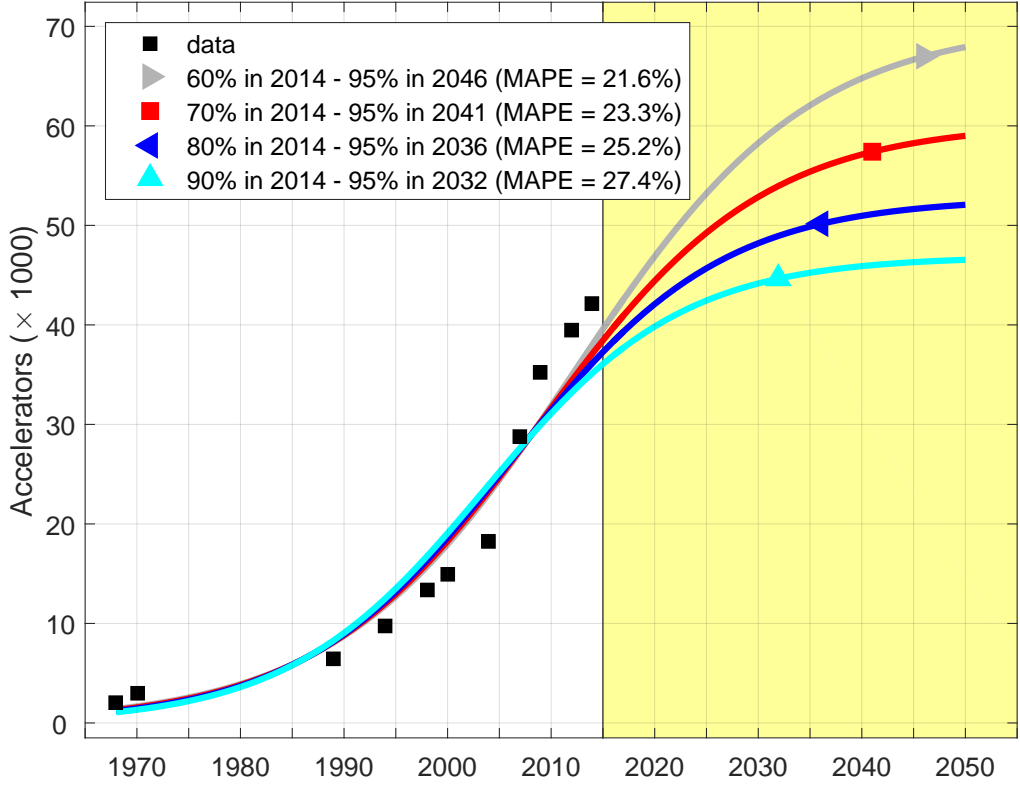

Figure 2: Number of accelerators in the world: scenarios for the 2015-2050 period

*Notes:* authors' calculation based on data from Table 1 in [7]. Lines represent fitted values from a nonlinear regression based on the logistic function of the total number of accelerators on a constant and a time trend. The legend shows for each of the four lines the percent of total demand that is assumed to be satisfied in 2014 and the year when such percent reaches 95%. The legend also shows the Mean Absolute Percent Error (MAPE) calculated over the 1968–2014 period; in–sample goodness of fit is inversely proportional to the MAPE of a model and is defined as: $MAPE = 100 \times n^{-1} \sum_t^n |(x_t - \hat{x}_t)/x_t|$, where $n$, $x_t$ and $\hat{x}_t$ denote the number of in–sample observations, the actual and fitted values, respectively. The shaded area identifies the out–of–sample time span, that is 2015-2050

As a preliminary step to extrapolate the future trajectories of the accelerator business, Figure 1 plots the data provided by [7] on the cumulated count of accelerators worldwide by field of application. As we can see, the average yearly growth rate is respectively 1%, 9%, 6% for science, medicine and industry, thus suggesting a great variability across fields of application. Accelerators in science feature a growth process that is well approximated by linear function. On the contrary, the growth dynamics of accelerators in industry and medicine is highly nonlinear.

As a second step, we suggest that the dynamics of the total number of accelerators worldwide is well approximated by an "$S$–shaped" time trend. Curves with this shape have a long tradition in statistics and are widely used in biology, demography and economics [19, 23]. In innovation studies, $S$–shaped curves describe how the adoption of a new technology evolves over time: diffusion rates first increase and then drop over time, giving rise to a period of very rapid diffusion, preceded by a slow take–off and followed in late periods by slow convergence to satiation and decline [34, 35]. The logistic function, is a leading example of an $S$–shaped time trend used to represent such pattern.[2]

Earlier literature motivates why the cumulative diffusion of technology is often $S$–shaped [23]. The "competition–legitimation" theory posits that some technologies become increasingly accepted as the number of users grows ("legitimation"), but then competition for limited resources acts as a cap on the maximum level of diffusion. Heterogeneity of population and countries is a further possible explanation. The income heterogeneity hypothesis posits that as the price of an innovation falls, more consumers can afford it. In this case $S$–curves emerge if the income distribution is bell–shaped and the price decreases monotonically. Information cascades are yet another explanation: the initial slow pace in the diffusion of a technology is due to the fact that early adopters start experimenting with different competing alternatives and then, once they have decided which to use, followers copy the choice of the precursors. Cross–country diffusion models that lead to an $S$–shaped curve are also available [28]. Potential adopters in lagging countries observe the introduction and diffusion of technologies in leader countries, that therefore bear much of the risk of the innovation. Once this risk has been reduced the technology quickly spreads in lagging countries and then slows down because of constraints in adoption capacity. See [9, 23, 30] for a survey.

Figure 2 relies on the logistic function to fit an $S$–shaped time trend to the data on the total number of accelerators and forecast possible trajectories up to 2050. Since the upper asymptote of the logistic function represents the steady–state size of the market, which is unknown, instead of estimating a ceiling parameter, we have drawn four curves that differ in the assumption regarding such value. If the number of accelerators available in 2014 represents 60% of potential demand, then the 95% of market would be supplied by 2046. On the other hand, if the 2014 value is equivalent to 90% of potential demand, 95% of the demand would be satisfied by 2032. Visual inspection suggests that the industry will reach maturity in the next 15–30 years. The Mean Absolute Prediction Error (MAPE) shown in the legend of Figure 2 implies that the best in–sample goodness–of–fit is associated with the assumption that 60% of demand was satisfied in 2014. Interestingly, in terms of out–of–sample goodness–of–fit, this parametrization outperforms also models based on a linear or on an exponential trend function. In fact, we have computed the out–of–sample MAPE using the most recent observations (2007–2014) and re-estimating the models excluding these data points. In this case, the MAPE is respectively 12%, 25%, 16% for the logistic, linear and exponential trend model. Inspection of Figure 2 suggests the existence of an inflection point in the accelerators' count dynamics in recent years.

We know from earlier literature [40] that a technological breakthrough is likely at this stage, given that it is the only way to change the convergence to a steady state. Figure 2 is somehow related with the so–called Livingston plot that shows how the laboratory energy of the particle beams produced by accelerators has nonlinearly increased through time [32]. The Livingstone plot is a useful and well appreciated device which illustrates that the key driver of accelerators' energy increase is a succession of new technologies, rather than the improvement or further development of existing machines and their diffusion [28].

---

[2]The logistic function arises as a solution of a first order non–linear differential equation of the form $\frac{d}{dx}f(x) = f(x)[1-f(x)]$. It implies that the growth over time of $x$ (e.g. a population or the stock of innovations) is a self–limiting process. Its use in economics dates back at least to [24]; [23] details its use in models of technology diffusion. While there are different parametrizations of the logistic function, we rely on the following. Let $y(t)$ be the number of users of a technology at time $t$, let $N$ be the size of the market (i.e. the maximum number of users) and let $t_0$ be the number of users when half of the total demand has been satisfied, then the logistic function can be written as: $y(t) = \frac{1}{1+e^{-\delta(t-t_0)}}$. The parameter $\delta > 0$ determines the steepness of the resulting $S$–shaped curve. Estimation implemented in Matlab with the `glmfit` routine.

The question is: *"how can we analyse the impact of such likely, albeit highly uncertain, technological innovation?"*

## 3 Social cost–benefit analysis and two case studies

Social Cost–Benefit Analysis (CBA, hereafter) can provide a conceptual framework to quantify and monetize the impact of a technological breakthrough. CBA has been routinely used by governments and international organizations to evaluate the net benefits of investment projects [18, 27]. More recently, the European Commission has included for the first time a chapter on the evaluation of research infrastructures (RI) in its reference guide for funding major projects under the EU Structural and Investment Funds, now in its fifth edition. See [16]. In fact, a positive social CBA evaluation is required for co–financing major projects with the European Regional Development Fund and the Cohesion Fund.[3] Moreover, *"Horizon 2020 – Work Programme 2018–2020. European research infrastructures"* [17] mentions that the preparatory phase of new ESFRI projects (www.esfri.eu) should include a CBA and refers to [16]. A CBA model for the evaluation of RI, recently proposed by [21], can be written as:

$$\begin{aligned} \mathbb{E}\left(NPV_{RI}\right) &= \mathbb{E}\left[NPV_u + PV_{B_n}\right] \\ &= \mathbb{E}\left[(PV_{B_u} - PV_{EC}) + PV_{B_n}\right] \end{aligned} \quad (1)$$

where $NPV_{RI}$ is the Net Present Value (NPV) of the RI that is defined as the sum of the NPV for users of the infrastructure ($NPV_u$, or net use–benefits) and the present value for "non–users" ($PV_{B_n}$, or non–use benefits). The latter term encompasses both possible benefits that might derive in the future from the RI and the "public good value" of new scientific knowledge. Equation (1) exploits the fact that net use–benefits are defined as the difference between the present economic value of benefits for users of the RI ($PV_{B_u}$) and the present value of its economic costs ($PV_{EC}$). All variables are "present values" in that their future and past values have been converted in the same unit of measure — the monetary value in a base year — using a given social discount rate and aggregated over the time horizon of the CBA [18, 27, for details]. Lastly, we note that all variables are stochastic and hence $\mathbb{E}(\cdot)$ denotes the expect value conditional on their probability density function.[4]

---

[3]According to Article 100 of EU Regulation No 1303/2013, a major project is an investment operation comprising *"a series of works, activities or services intended to accomplish an indivisible task of a precise economic and technical nature which has clearly identified goals and for which the total eligible cost exceeds EUR 50 million."*

[4]Since variables might not be independent, we cannot simplify the equation further and write the right–hand side as the sum of expected values.

Figure 3: Social CBA of the LHC & the National Centre of Oncological Hadrontherapy

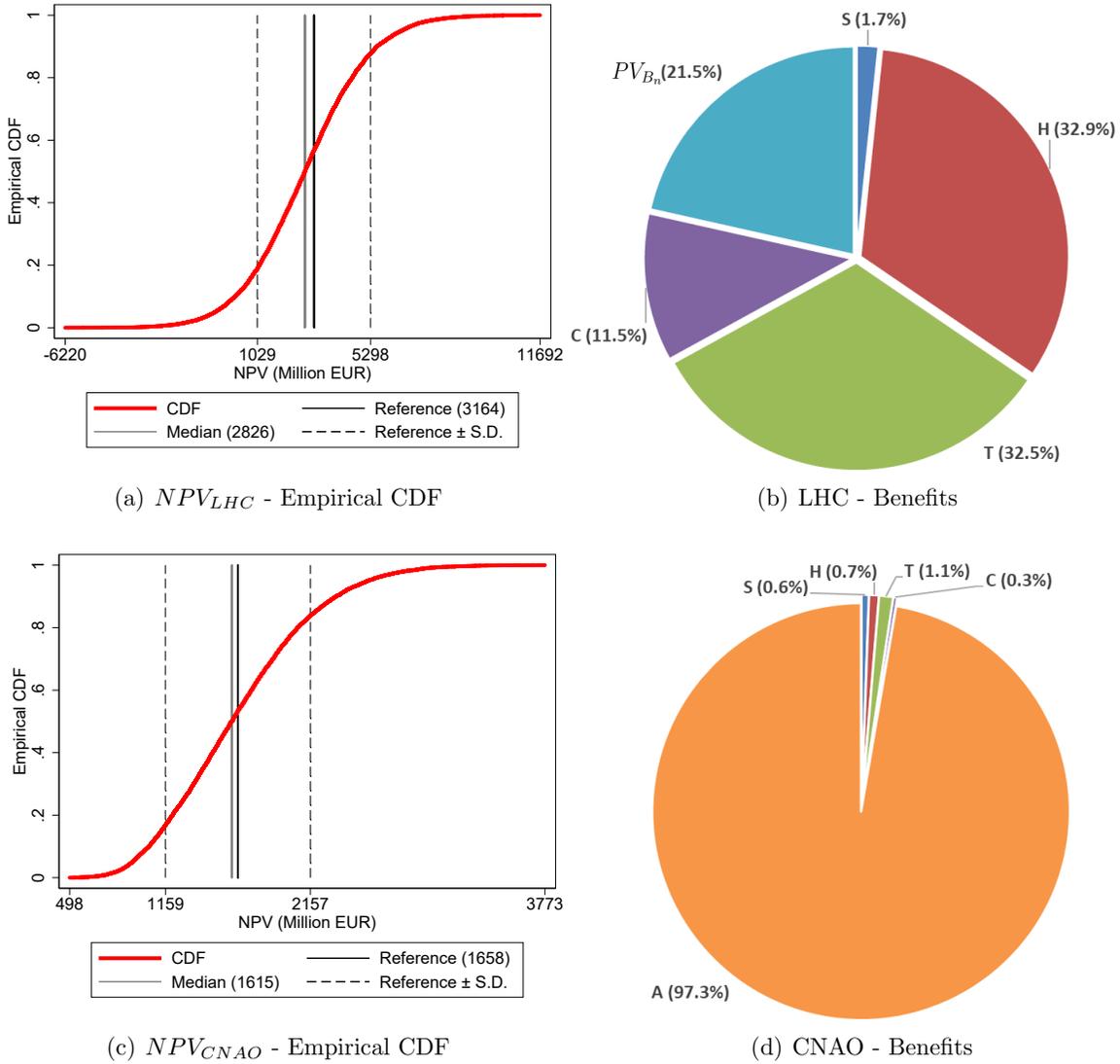

(a) $NPV_{LHC}$ - Empirical CDF

(b) LHC - Benefits

(c) $NPV_{CNAO}$ - Empirical CDF

(d) CNAO - Benefits

*Notes:* Panel (a) and (b) adapted from [20], while panel (c) and (d) are adapted from [3]. Panel (a) and (c) show the empirical cumulative density function of the NPV of LHC and CNAO, based on Monte Carlo simulations. Figures in panel (b) and (d) show the share of the benefits for each category of stakeholder, namely: the value of academic publications ($S$), technological spillovers ($T$), human capital ($H$), cultural effects ($C$), other users' benefits of applied research ($A$) and non–use benefits ($PV_{B_n}$). In Panel (b) $PV_{B_n}$ represents the existence value of LHC, while in Panel (d) $A$ encompasses both health benefits for patients (95.1%) and benefits for users of the experimental beam (2.2%).

Net use–benefits of a RI are the sum of benefits to different stakeholders: the value of academic publications for scientists ($S$), technological spillovers for collaborating firms and other economic agents ($T$), human capital accumulation for early career researchers ($H$), cultural effects for the general public ($C$) and other users' benefits of applied research ($A$). Similarly, economic costs can be broken–down into investment and operating costs ($TC$) and any negative externality ($E$, e.g. pollution). It is thus apparent that a social CBA is different from the financial appraisal of a project. Therefore, equation (1) becomes:

$$\mathbb{E}\left(NPV_{RI}\right) = \mathbb{E}\left[(S + T + H + C + A) - (TC + E) + PV_{B_n}\right] \quad (2)$$

This model has been implemented in two recent studies assessing the social benefits of accelerators in science and medicine: the Large Hadron Collider (LHC) at CERN [20] and the National Centre of Oncological Hadrontherapy (CNAO) in Pavia, Italy [3]. A summary of these analyses appears in Figure 3. Both projects yield $\mathbb{E}(NPV_{RI}) > 0$ and hence successfully pass the CBA test, albeit with different probabilities. The composition of total benefits is also different. For CERN, 33% of the total benefit is due to improved career opportunities for former students and researchers ($H$) and 33% of the total is associated with technological spillovers ($T$), while the lion's share of CNAO's benefits is not surprisingly due to health benefits for patients ($A$), that represent 97% of the total.

## 4 From individual accelerators to the industry

We now move from a CBA model for a single RI to a model for the industry. Figure 4 represents a sketch of the market for accelerators. In each panel the horizontal axis represents the (standardized) quantity of accelerators produced and purchased per year, while the marginal willingness to pay of buyers (e.g. hospitals, industry, universities etc) and the price required by producers appears on the vertical axis. The latter reflects marginal costs including R&D for the new technologies. The starting point is panel (a) where the market is at equilibrium point $A_0$.

When the equilibrium price is $p_0$ and the equilibrium quantity is $q_0$, we identify the consumer surplus with the hatched area and the producer surplus with the shaded area, while we use term "economic surplus" to indicate their sum [4]. The consumer surplus is given by the difference between the maximum price a consumer is willing to pay and the equilibrium price $p_0$. It can be represented as:

$$CS_0 = \int_0^{q_0} D_0(q)\, dq - p_0 q_0 \tag{3}$$

Similarly, the producer surplus is the additional private benefit to firms, in terms of profits, when the market price is greater than the minimum price at which they are willing to sell:

$$PS_0 = p_0 q_0 - \int_0^{q_0} S_0(q)\, dq \tag{4}$$

Figure 4(b) shows the impact of a technological breakthrough due to the appearance of a new generation of accelerators produced at a lower unit cost (upon a standardized measure of their performance) and hence sold at a lower price: the new equilibrium is at point $A_1$, with a lower price $p_1$ and a higher quantity $q_1$. The hatched and shaded areas now represent the change in consumer and producer surplus, respectively: a technological breakthrough yields an increase in economic surplus for the whole society. These first–round effects are however just a share of the total increase in benefits due to the new technology. In fact, there are some additional second–round effects to third parties because more accelerators are now available ($\Delta q$).

Figure 4: Socio-economic impacts of a breakthrough technological change in the technology of accelerators

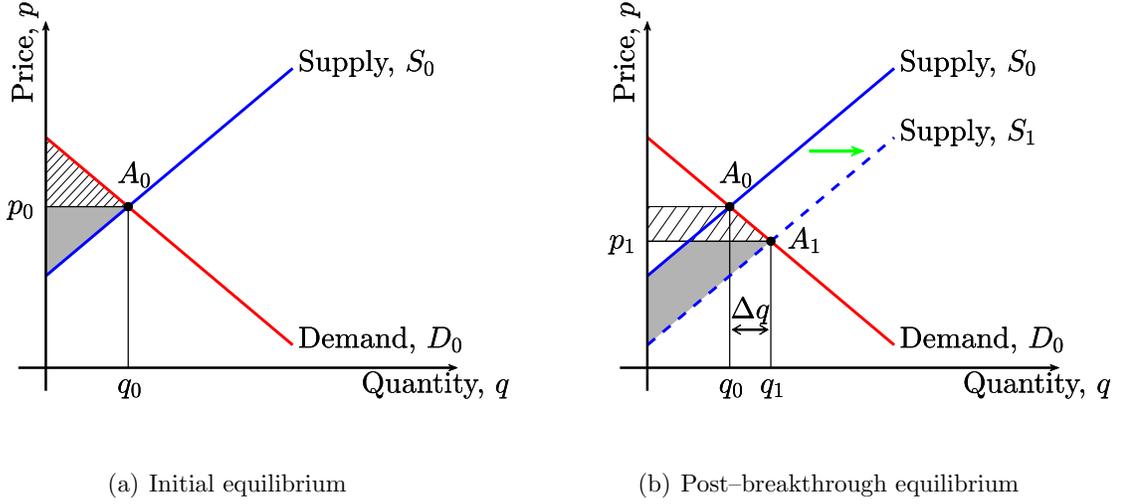

(a) Initial equilibrium            (b) Post–breakthrough equilibrium

*Notes:* panel (a) shows the market equilibrium — given by the intersection of the demand ($D_0$) and supply ($S_0$) curves — consumer surplus ($CS$, hatched area) and producer surplus ($PS$, shaded area) before the technological breakthrough. Panel (b) shows the increase in consumer surplus ($\Delta CS$, hatched area) and producer surplus ($\Delta PS$, shaded area) after the breakthrough. $\Delta q$ is the quantity effect that in turns generates the indirect effects in equation (5). These are: the value of academic publications for scientists ($\Delta S$), technological spillovers for collaborating firms and other economic agents ($\Delta T$), human capital accumulation for early career researchers ($\Delta H$), cultural effects for the general public ($\Delta C$) and other users' benefits of applied research ($\Delta A$).

This leads to the following equation for a technology breakthrough (TB):

$$\begin{aligned} \mathbb{E}\left(NPV_{TB}\right) &= \mathbb{E}\left(\Delta CS + \Delta PS + \Delta PV_{B_u} - PV_R\right) \\ &= \mathbb{E}\left(\Delta CS + \Delta PS + \Delta S + \Delta H + \Delta T + \Delta C + \Delta A - PV_R\right) \end{aligned} \qquad (5)$$

where $\Delta$ is the difference operator, such that for a generic variable $Y$: $\Delta Y = Y_1 - Y_0$. Notice that we now omit the non–use benefit ($PV_{B_n}$) because it is related to new scientific discoveries, not to technological progress, but subtract the present value of pre–commercial research expenditure ($PV_R$), financed mostly with public funds. This is the total amount of funds supporting scientific and technological research on advanced accelerator concepts, before firms start allocating part of their R&D expenditure to produce accelerators for the market.

# 5 Discussion and a research agenda

A possible research agenda on the socio–economic impacts of a breakthrough in accelerator technology is based on the model in Equation (5) and involves three steps: ($i$) a long-term forecast of the global demand for particle accelerators; ($ii$) a scenario analysis of the technological change; ($iii$) the assessment of the potential socio–economic benefits of the transition to new technologies to different stakeholders. We briefly discuss each of these steps.

(*i*) *Demand scenario analysis.* More detailed data about supply– and demand–side drivers can be used to improve the scenarios in Figure 2; this entails breaking down the study into areas of application. Two main demand curves should be predicted: that for accelerators for medical applications and that for accelerators for industrial and other non–medical services[5]. A finer grain analysis of the socio–economic effects should focus also on some key segments, such as ion implantation on semiconductor devices or cancer therapy. Then price elasticities should be estimated to make the model sketched in Figure 4 operational: this would allow to estimate different demand curves and hence first round effects of price changes triggered by a technological change.

(*ii*) *Technological forecasting analysis.* [8] report that in the US a series of consultations among key proponents of different advanced accelerator concepts has delivered broad–view roadmaps to an $e^+e^-$ collider operating by about 2040 and more detailed roadmaps covering intermediate steps over the next decade. Different roadmaps may be available elsewhere. A judgmental forecasting exercise seems thus appropriate in this context: this approach is based on gathering an international panel of experts and elicit their views about likely future scenarios.[6] This information is used to estimate the probability distribution function of different technological scenarios, conditional to existing information [6, 31]. The Delphi method involves multi–round forecasting challenges where experts provide initial forecasts and then adjust their initial guesses based on feedbacks they receive.[7] This process is iterated until a satisfactory level of consensus is reached and final forecasts are constructed from the aggregation of individual forecasts. See [10] for an application of the Delphi method to the design a technology roadmap.

(*iii*) *Assessing socio–economic net benefits.* Combining demand and technological scenarios (*i*) and (*ii*), a CBA can be implemented through the estimation of Equation (5). Potential net socio–economic benefits are driven by the difference between the cost trajectory of the existing technologies and the cost trajectory of the new technologies, plus the benefits beyond direct incremental effects for producers and consumers. These are:

- Effects on scientific knowledge creation ($\Delta S$): the increased availability at a lower cost of experimental data; the counterfactual scenario here is the investment and operating costs needed to generate the same data with the existing technologies or their development along the same trajectory.

- Effects on human capital ($\Delta H$): PhD students and post–doc fellows in laboratories equipped with the new technology not only benefit from training, but might also improve their career opportunities and hence their long–term salary; the counterfactual is given by a lower number of students able to access research accelerators, particularly in less developed economies.

- Learning spillovers for firms ($\Delta T$): arising because of the need to solve new problems, potentially creating new products and processes, or learning–by–doing effects because of the increased production scale.

---

[5]We assume that the demand for accelerators for science is relatively inelastic to price.

[6]See [39] for an alternative approach.

[7]The Delphi method was developed at the RAND Corporation in the 1950's for predicting bombing requirements in a cold–war conflict with the USSR [12]. Surveys of the Delphi method are provided by [13], [36] and [37], while [29] review the judgmental forecasting approach.

- Cultural effects ($\Delta C$): due to the easier outreach made possible by smaller, less expensive, experimental equipment and possibly some effects on the general public.

- Incremental benefits to final consumers ($\Delta A$): for example, for patients because of increased availability of radiotherapy in less developed economies.

These benefits are quantitatively represented by the expected net present value of the difference between the two combined demand–technological scenarios (current trajectory versus breakthrough), over a suitable long–term horizon, perhaps around 50 years from now, given a social discount rate. The [16] suggests a 3% social discount rate, which implies that a benefit of one Euro occurring at year 50 is valued just $1/(1+0.03)^{50} \approx 0.23$ Euro. Given the uncertainty surrounding both the demand drivers and the cost reductions over such a long-time span, several variables in the forecasting model should be treated as stochastic and the final result should be expressed in the form of a conditional probability distribution of the NPV. A further complication is due to the uncertainty surrounding the time horizon of the analysis [21].

Our analysis suggests that the industry has entered the declining growth rate portion of an $S$–shaped trajectory. This is a symptom of maturity of the technology, as observed in many other sectors. In this situation, reliable early detection of a possible technological breakthrough in acceleration science, and the forecast of the highly uncertain benefits it could generate, is key for attracting governmental funds to research.

# Acknowledgements

The authors are grateful for comments on a previous version to an anonymous referee, to Eric Colby (Office of High Energy Physics, US Department of Energy), Massimo Ferraio (INFN, National Institute for Nuclear Physics), Chiara Pancotti (CSIL, Centre for Industrial Study, Milan) and several participants at the 3rd European Advanced Accelerator Concepts Workshop. The research has been funded by the University of Milan. The usual disclaimer applies.